\documentclass[fleqn]{article}

\usepackage{espcrc2,graphicx}

\renewcommand{\theequation}{\arabic{section}.\arabic{equation}}

\begin{document}

\title{
\vspace*{-1cm}
\begin{flushright}
\begin{minipage}{3cm}
\begin{flushleft}
 {\normalsize DPNU-03-15}
\end{flushleft}
\end{minipage}
\end{flushright}
\vspace*{-0.3cm}
\bf Non-renormalization Theorem \\
    Originating in 
    a New Fixed Point of the Vector Manifestation}
\author{
{\bf Chihiro Sasaki}\address[Nagoya]{%
Department of Physics, Nagoya University,
Nagoya, 464-8602, Japan.}
}

\begin{abstract}
We study the pion velocity at the critical temperature of chiral
symmetry restoration in QCD. Starting from the premise that the
bare effective field theory is to be defined from the underlying
QCD, we incorporate the effects of Lorentz non-invariance into the
bare theory by matching an effective field theory to QCD at a
suitable matching scale, 
and investigate how the Lorentz non-invariance existing 
in the bare theory influences physical quantities.
Using the hidden local symmetry model as
the effective field theory, where the chiral symmetry restoration
is realized as the vector manifestation (VM), we find that the
pion velocity at the critical temperature receives neither quantum
nor (thermal) hadronic corrections at the critical temperature
even when we start from the bare theory with Lorentz symmetry
breaking. This is likely the manifestation of a new fixed point
in the Lorentz non-invariant formulation of the VM.
\end{abstract}

\maketitle

\section{Introduction}

Chiral symmetry in QCD is expected to be restored under some
extreme conditions such as large number of flavor $N_f$ and high
temperature and/or density. In hadronic sector, the chiral
symmetry restoration is described by various effective field
theories (EFTs)
based on the chiral symmetry ~\cite{review:chiral}. 

By using the hidden local symmetry (HLS)
model~\cite{Bando} as an EFT
and performing the Wilsonian matching which is one of the methods
that determine the bare theory from the underlying QCD~\cite{HY:WM}, 
the vector manifestation (VM) in hot
or dense matter was formulated in Refs.~\cite{HS:VM,HKR:VM}. In the VM,
the massless vector meson becomes the chiral partner of pion at
the critical point~\cite{HY:VM}~\footnote{
 As studied in Ref.~\cite{HY:PR} in detail,
 the VM is defined only as a limit with bare parameters 
 approaching the VM fixed point from the broken phase.
}.
There, the {\it intrinsic temperature or
density dependences of the parameters} of the HLS Lagrangian,
which are obtained by integrating out the high energy modes
(i.e., the quarks and gluons above the matching scale) in hot
and/or dense matter, play
the essential roles to realize the chiral symmetry restoration
consistently.

In the analysis done in
Ref.~\cite{HKRS:SUS}, it was shown that the effect of Lorentz
symmetry breaking to the bare parameters caused by the intrinsic
temperature dependence through the Wilsonian matching are
small~\cite{HKRS:SUS,HS:VVD}. 
Starting from the bare Lagrangian with Lorentz invariance, 
it was presented that the pion velocity
approaches the speed of light at the critical temperature~\cite{HKRS:SUS}, 
although in low temperature region $(T \ll
T_c)$ the pion velocity deviates from the speed of light due to
hadronic corrections~\cite{HS:VVD}.

However there do exist the Lorentz non-invariant effects 
in bare EFT anyway
due to the intrinsic temperaure and/or density effects.
Further the Lorentz non-invariance might be enhanced through the 
renormalization group equations (RGEs),
even if effects of Lorentz symmetry breaking at the bare level are small.
Thus it is important to investigate how the Lorentz non-invariance 
at bare level influences physical quantities.  

In this paper, 
we pick up the pion velocity at the critical temperature
and study the quantum and hadronic thermal effects
based on the VM.
The pion velocity is one of the important quantities since it controls
the pion propagation in medium through the dispersion relation.

Our main result is that {\it the pion velocity does not
receive either quantum or hadronic corrections 
in the limit $T \to T_c\,$}:
\begin{equation}
\qquad
 v_\pi(T) = V_{\pi,{\rm bare}}(T),
\label{result}
\end{equation}
independently of the value of the bare pion velocity
$V_{\pi,{\rm bare}}$.
This non-renormalization property on the pion velocity
is protected by the VM.
Equation~(\ref{result}) implies that the Lorentz non-invariance 
at bare level
is directly reflected on the physical pion velocity at
the critical temperature.

This paper is organized as follows:
In section~\ref{HLS}, we show the HLS Lagrangian with
Lorentz non-invariance and the conditions satisfied at the critical
temperature for the bare parameters.
In section~\ref{VMFP}, we show that the conditions for the bare
parametes at the critical temperature are satisfied in any energy scale
and that this is protected as a fixed point of the relevent RGEs.
In section~\ref{PVCT}, we show the quantum and hadronic corrections
to the pion velocity and derive our result~(\ref{result}).
In section~\ref{SD}, we give a summary and discussions.

\setcounter{equation}{0}
\section{Hidden Local Symmetry}
\label{HLS}


Since Lorentz symmetry breaking effects are included in the bare
theory through the Wilsonian matching, the HLS Lagrangian in hot
and/or dense matter is generically Lorentz non-invariant. Its
explicit form was presented in Ref.~\cite{HKR:VM}. In this
section, we start from this Lagrangian with Lorentz
non-invariance, and requiring that the axial-vector current correlator be
equal to the vector current correlator at the critical point, we
obtain the conditions for the bare parameters.


\subsection{Lorentz Non-invariant HLS Lagrangian}
\label{LNI-HLSL}

In this subsection, we show the HLS Lagrangian at leading order
including the effects of
Lorentz non-invariance.

The HLS model is based on
the $G_{\rm{global}} \times H_{\rm{local}}$ symmetry,
where $G=SU(N_f)_L \times SU(N_f)_R$ is the chiral symmetry
and $H=SU(N_f)_V$ is the HLS.
The basic quantities are
the HLS gauge boson $V_\mu$ and two matrix valued
variables $\xi_L(x)$ and $\xi_R(x)$
which transform as
$\xi_{L,R}(x) \to \xi^{\prime}_{L,R}(x)
  =h(x)\xi_{L,R}(x)g^{\dagger}_{L,R}$,
where $h(x)\in H_{\rm{local}}\ \mbox{and}\ g_{L,R}\in
[\mbox{SU}(N_f)_{\rm L,R}]_{\rm{global}}$.
These variables are parameterized as
\footnote{
 The wave function renormalization constant of the pion field
 is given by the temporal component of the pion decay constant
 ~\cite{Meissner:2001gz}.
 Thus we normalize $\pi$ and $\sigma$ by $F_\pi^t$ and $F_\sigma^t$
 respectively.
}
\begin{equation}
\qquad
  \xi_{L,R}(x)=e^{i\sigma (x)/{F_\sigma^t}}
     e^{\mp i\pi (x)/{F_\pi^t}},
\end{equation}
where $\pi = \pi^a T_a$ denotes the pseudoscalar Nambu-Goldstone
bosons associated with the spontaneous symmetry breaking of
$G_{\rm{global}}$ chiral symmetry, and $\sigma = \sigma^a T_a$
denotes the Nambu-Goldstone bosons associated with the spontaneous
breaking of $H_{\rm{local}}$. This $\sigma$ is absorbed into the
HLS gauge boson through the Higgs mechanism, and then the vector
meson acquires its mass. $F_\pi^t$ and $F_\sigma^t$ denote
the temporal components of the decay constant of
$\pi$ and $\sigma$, respectively.
The covariant derivative of $\xi_{L}$ is given
by
\begin{equation}
\qquad
 D_\mu \xi_L = \partial_\mu\xi_L - iV_\mu \xi_L + i\xi_L{\cal{L}}_\mu,
\end{equation}
and the covariant derivative of $\xi_R$ is obtained
by the replacement of ${\cal L}_\mu$ with ${\cal R}_\mu$
in the above where
$V_\mu$ is the gauge field of $H_{\rm{local}}$, and
${\cal{L}}_\mu$ and ${\cal{R}}_\mu$ are the external
gauge fields introduced by gauging $G_{\rm{global}}$ symmetry.
In terms of ${\cal L}_\mu$ and ${\cal R}_\mu$,
we define the external axial-vector and vector fields as
${\cal A}_\mu = ( {\cal R}_\mu - {\cal L}_\mu )/2$ and
${\cal V}_\mu = ( {\cal R}_\mu + {\cal L}_\mu )/2$.

In the HLS model it is possible to perform the derivative
expansion systematically~\cite{Georgi,Tanabashi:1993sr,HY:PR}. 
In the chiral perturbation theory (ChPT) with  HLS, 
the vector meson mass is to be considered as small compared with 
the chiral symmetry breaking scale $\Lambda_\chi$, 
by assigning ${\cal O}(p)$ to the HLS gauge coupling, 
$g \sim {\cal O}(p)$~\cite{Georgi,Tanabashi:1993sr}. 
(For details of the ChPT with HLS, see Ref.~\cite{HY:PR}.) 
The leading order Lagrangian with
Lorentz non-invariance can be written as~\cite{HKR:VM}
\begin{eqnarray}
&&
{\cal L}
=
\biggl[
  (F_{\pi}^t)^2 u_\mu u_\nu
  {}+ F_{\pi}^t F_{\pi}^s
    \left( g_{\mu\nu} - u_\mu u_\nu \right)
\biggr] \nonumber\\
&&\qquad\qquad\times
\mbox{tr}
\left[
  \hat{\alpha}_\perp^\mu \hat{\alpha}_\perp^\nu
\right] \nonumber\\
&&{}+
\biggl[
  (F_{\sigma}^t)^2 u_\mu u_\nu
  {}+  F_{\sigma}^t F_{\sigma}^s
    \left( g_{\mu\nu} - u_\mu u_\nu \right)
\biggr] \nonumber\\
&&\qquad\qquad\times
\mbox{tr}
\left[
  \hat{\alpha}_\parallel^\mu \hat{\alpha}_\parallel^\nu
\right]
\nonumber\\
&&
{} +
\Biggl[
  - \frac{1}{ g_{L}^2 } \, u_\mu u_\alpha g_{\nu\beta}
  {}- \frac{1}{ 2 g_{T}^2 }
  \left(
    g_{\mu\alpha} g_{\nu\beta}
   - 2 u_\mu u_\alpha g_{\nu\beta}
  \right)
\Biggr] \nonumber\\
&&\qquad\qquad \times
\mbox{tr}
\left[ V^{\mu\nu} V^{\alpha\beta} \right]
\ ,
\label{Lag}
\end{eqnarray}
where
\begin{equation}
\qquad
 \hat{\alpha}_{\perp,\parallel }^{\mu}
 = \frac{1}{2i}\bigl[ D^\mu\xi_R \cdot \xi_R^{\dagger}
                 {}\mp  D^\mu\xi_L \cdot \xi_L^{\dagger}
                   \bigr].
\end{equation}
Here $F_{\pi}^s$ denote the spatial pion decay constant and similarly
$F_{\sigma}^s$ for the $\sigma$. The rest frame
of the medium is specified by $u^\mu = (1,\vec{0})$ and
$V_{\mu\nu}$ is the field strength of $V_\mu$. $g_{L}$ and $g_{T}$
correspond in medium to the HLS gauge coupling $g$. The parametric
$\pi$ and $\sigma$ velocities are defined by~\cite{Pisarski:1996mt}
\begin{equation}
\qquad
 V_\pi^2 = {F_\pi^s}/{F_\pi^t}, \qquad
 V_\sigma^2 = {F_\sigma^s}/{F_\sigma^t}.
\end{equation}


\subsection{Conditions for the Bare Parameters}
\label{VMC}

In this subsection following Ref.~\cite{HKR:VM} where the conditions
for the current correlators
with the bare parameters in dense matter were presented,
we show the Lorentz non-invariant version of the conditions satisfied 
at the critical temperature for the bare parameters.

Concept of the matching in the Wilsonian sense is based on the following
assumptions:
The bare Lagrangian of the effective field theory (EFT) 
${\cal L}_{\rm bare}$ is defined at a suitable matching scale $\Lambda$.
Generating functional derived from ${\cal L}_{\rm bare}$ leads to
the same Green's function as that derived from the generating functional
of QCD at $\Lambda$.
In other words, the bare parameters are obtained after integrating out
the high energy modes, i.e., the quarks and gluons above $\Lambda$.
When we integrate out the high energy modes in hot matter,
the bare parameters have a certain 
temperature dependence, {\it intrinsic temperature dependence}, 
converted from QCD to the EFT.
The intrinsic temperature dependence is nothing but the signature
that hadrons have an internal structure constructed from 
quarks and gluons.
In the following, we describe the chiral symmetry restoration
based on the point of view that {\it the bare HLS theory is
defined from the underlying QCD.}
We note that the Lorentz non-invariance appears in the bare HLS theory
as a result of including the intrinsic temperature dependence.
Once the temperature dependence of the bare parameters is determined 
through the matching with QCD mentioned above,
the parameters 
appearing in the hadronic corrections pick up the intrinsic 
thermal effects through the RGEs.

The axial-vector and vector current correlators at bare level
are constructed in terms of bare parameters and
are divided into the longitudinal and transverse components:
\begin{equation}
\qquad
 G_{A,V}^{\mu\nu} = P_L^{\mu\nu}G_{A,V}^L + P_T^{\mu\nu}G_{A,V}^T,
\end{equation}
where $P_{L,T}^{\mu\nu}$ are the longitudinal and transverse 
projection operators, respectively.
The axial-vector current correlator 
in the HLS around the matching scale $\Lambda$ is well described 
by the following forms with the bare parameters~\cite{HKR:VM,HKRS:SUS}:
\begin{eqnarray}
 &&
 G_{A{\rm(HLS)}}^L(p_0,\bar{p})
 =
 \frac{ p^2 F_{\pi,{\rm bare}}^t F_{\pi,_{\rm bare}}^s }{
    - [ p_0^2 - V_{\pi,{\rm bare}}^2 \bar{p}^2 ] }
 -2p^2z^L_{2,\rm bare},
\label{gal} \nonumber\\
 &&
 G_{A{\rm(HLS)}}^T(p_0,\bar{p})
 =
 -F_{\pi,\rm bare}^tF_{\pi,\rm bare}^s
\nonumber\\
 &&\qquad\qquad\qquad
 {} - 2 \left(
    p_0^2 z_{2,{\rm bare}}^L  - \bar{p}^2 z_{2,{\rm bare}}^T
  \right)
\ ,
\label{gat}
\end{eqnarray}
where $z_{2,{\rm bare}}^{L}$ and $z_{2,{\rm bare}}^{T}$ are
the coefficients of the higher order terms,
and $V_{\pi,{\rm bare}}$ is the bare pion velocity related to
$F_{\pi,{\rm bare}}^t$ and $F_{\pi,{\rm bare}}^s$ as
\begin{equation}
\qquad
V_{\pi,{\rm bare}}^2
= \frac{F_{\pi, {\rm bare}}^s}{F_{\pi, {\rm bare}}^t}
\ .
\end{equation}
Similarly, two components of 
the vector current correlator have the following forms:
\begin{eqnarray}
&&G_{V{\rm(HLS)}}^L(p_0,\bar{p})
\nonumber\\
&&=
\frac{
  p^2 \, F_{\sigma,{\rm bare}}^t F_{\sigma,{\rm bare}}^s
  \left( 1 - 2 g_{L,{\rm bare}}^2 z_{3,{\rm bare}}^L \right)}
 { -
  \left[
    p_0^2 - V_{\sigma,{\rm bare}}^2 \bar{p}^2
    - M_{\rho,{\rm bare}}^2
  \right] } 
\nonumber\\
&&\quad
{}- 2 p^2 z_{1,{\rm bare}}^L
+ {\cal O}(p^4)
\ ,
\nonumber\\
&&G_{V{\rm(HLS)}}^T(p_0,\bar{p})=
\frac{
  F_{\sigma,{\rm bare}}^t F_{\sigma,{\rm bare}}^s}
 { - \left[
    p_0^2 - V_{T,{\rm bare}}^2 \bar{p}^2 - M_{\rho,{\rm bare}}^2
  \right] }
\nonumber \\
&&
\times
  \bigl[
    p_0^2 \left(1 - 2 g_{L,{\rm bare}}^2 z_{3,{\rm bare}}^L\right)
\nonumber\\
&&\qquad\qquad
   {} - V_{T,{\rm bare}}^2 \bar{p}^2
    \left(1 - 2 g_{T,{\rm bare}}^2 z_{3,{\rm bare}}^T\right)
  \bigr]
\nonumber\\
&&\quad
{}- 2 \left( p_0^2 z_{1,{\rm bare}}^L - \bar{p}^2 z_{1,{\rm bare}}^T \right)
+ {\cal O}(p^4)
\ ,
\end{eqnarray}
where $z_{1,2,{\rm bare}}^L$ and $z_{1,2,{\rm bare}}^T$ denote
the coefficients of the higher order terms.
In the above expressions, the bare vector meson mass in the rest frame, 
$M_{\rho,{\rm bare}}$, is
\begin{equation}
\qquad
 M_{\rho,{\rm bare}}^2
 \equiv g_{L,{\rm bare}}^2 F_{\sigma,{\rm bare}}^t
        F_{\sigma,{\rm bare}}^s \ .
\end{equation}
We define the bare parameters $a^t_{\rm bare}$ and $a^s_{\rm bare}$ as
\begin{equation}
\quad
  a^t_{\rm bare}
  = \Biggl( \frac{F_{\sigma,{\rm bare}}^t}
    {F_{\pi,{\rm bare}}^t} \Biggr)^2, \quad
  a^s_{\rm bare}
  = \Biggl( \frac{F_{\sigma,{\rm bare}}^s}
    {F_{\pi,{\rm bare}}^s} \Biggr)^2,
\end{equation}
and the bare $\sigma$ and transverse $\rho$ velocities as
\begin{equation}
\quad
 V_{\sigma,{\rm bare}}^2
 = \frac{F_{\sigma,{\rm bare}}^s}{F_{\sigma,{\rm bare}}^t}, \quad
 V_{T,{\rm bare}}^2
 = \frac{g_{L,{\rm bare}}^2}{g_{T,{\rm bare}}^2}.
\end{equation}

Now we consider the matching near the critical temperature.
At the chiral phase transition point, the axial-vector and vector
current correlators must agree with each other: $G_{A{\rm(HLS)}}^L
= G_{V{\rm(HLS)}}^L$ and $G_{A{\rm(HLS)}}^T = G_{V{\rm(HLS)}}^T$.
These equalities are satisfied for
any values of $p_0$ and $\bar{p}$ around the matching scale
only if the following conditions are met:
\begin{eqnarray}
&&
  a_{\rm bare}^t \to 1, \quad
  a_{\rm bare}^s \to 1,
\nonumber\\
&&
  g_{L,{\rm bare}} \to 0, \quad
  g_{T,{\rm bare}} \to 0  \quad \mbox{for}\,\,T \to T_c.
\end{eqnarray}
This implies that at bare level the longitudinal mode of the vector 
meson becomes the real NG boson and couples to the vector current correlator, 
while the transverse mode decouples.

\setcounter{equation}{0}
\section{Vector Manifestation Condition}
\label{VMFP}

In this section, we show that the conditions for the bare parameters
for $T \to T_c$ are satisfied in any energy scale and that
this is protected by the fixed point of the RGEs.

It was shown that the HLS gauge coupling $g=0$ is a fixed point of the RGE
for $g$ at one-loop level~\cite{HY:1-loop,HY:WM}.
The existence of the fixed point $g=0$ is guaranteed by
the gauge invariance.
This is easily understood from the fact that the gauge field is
normalized as $V_\mu = g \rho_\mu$.
In the present case without Lorentz symmetry,
the gauge field is normalized by $g_L$ as $V_\mu = g_L \rho_\mu$
and thus $g_L = 0$ becomes a fixed point of the RGE for $g_L$.

\begin{figure}
 \begin{center}
  \includegraphics[width = 6cm]{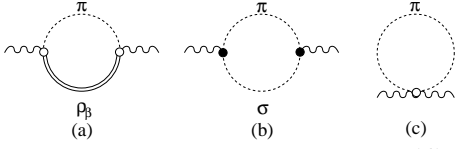}
 \end{center}
\vspace*{-1.4cm}
 \caption{Diagrams for contributions to $\Pi_\perp^{\mu\nu}$
         at one-loop level.}
 \label{fig:AAdiagrams}
\end{figure}

\begin{figure}
\vspace*{-0.5cm}
 \begin{center}
  \includegraphics[width = 7cm]{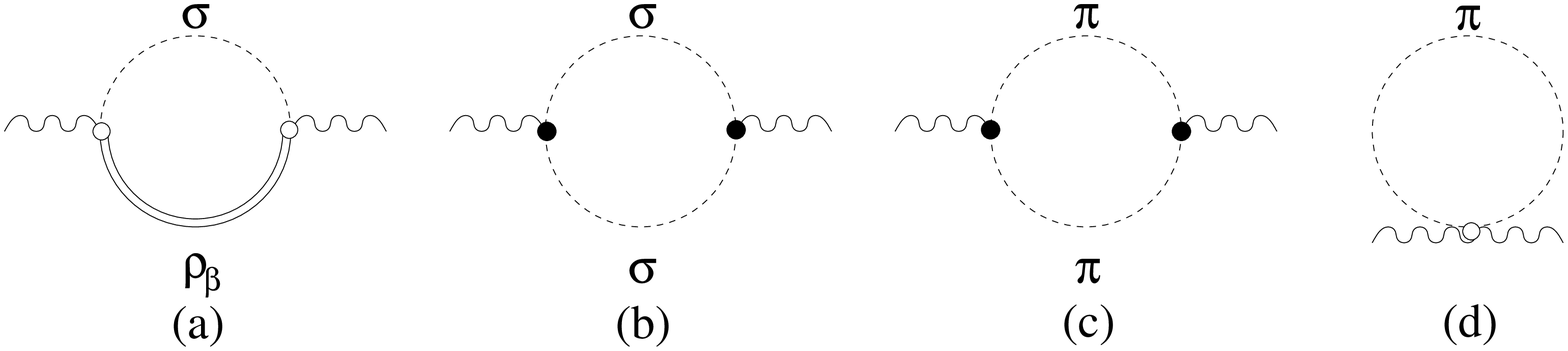}
 \end{center}
\vspace*{-1.4cm}
 \caption{Diagrams for contributions to
  $\Pi_\parallel^{\mu\nu}$ at one-loop level.}
 \label{fig:VVdiagrams}
\end{figure}

Provided that $g_L = 0$ is a fixed point,
we can show that $a^t = a^s = 1$ is also a fixed point 
of the coupled RGEs for $a^t$ and $a^s$ as follows:
We start from the bare theory defined at a scale $\Lambda$
with $a_{\rm bare}^t = a_{\rm bare}^s = 1$ (and $g_L = 0$).
The parameters $a^t$ and $a^s$ at $(\Lambda - \delta\Lambda)$
are calculated by integrating out the modes in
[$\Lambda - \delta\Lambda, \Lambda$].
They are obtained from the two-point functions of
${\cal A}_\mu$ and ${\cal V}_\mu$,
denoted by $\Pi_\perp^{\mu\nu}$ and $\Pi_\parallel^{\mu\nu}$.
We decompose these functions into
\begin{eqnarray}
&&
 \Pi_{\perp,\parallel}^{\mu\nu}
  =u^\mu u^\nu \Pi_{\perp,\parallel}^t +
   (g^{\mu\nu}-u^\mu u^\nu)\Pi_{\perp,\parallel}^s \nonumber\\
&&\qquad\qquad{}+
   P_L^{\mu\nu}\Pi_{\perp,\parallel}^L + 
   P_T^{\mu\nu}\Pi_{\perp,\parallel}^T,
\label{Pi perp decomp}
\end{eqnarray}
where $u^\mu u^\nu$, $(g^{\mu\nu}-u^\mu u^\nu)$, $ P_L^{\mu\nu}$
and $P_T^{\mu\nu}$ denote the temporal, spatial, longitudinal and
transverse projection operators, respectively.
The parameters $a^t$ and $a^s$ are defined by
$ a^t = {\Pi_\parallel^t}/{\Pi_\perp^t}\,,
  a^s = {\Pi_\parallel^s}/{\Pi_\perp^s}$~\cite{HS:VVD}.
These expressions are further reduced to
\begin{eqnarray}
 &&
 a^t(\Lambda - \delta\Lambda) = a_{\rm bare}^t
 {}+ \frac{1}{\bigl( F_{\pi,{\rm bare}}^t \bigr)^2}
\nonumber\\
 &&\times
   \Bigl[ \Pi_\parallel^t (\Lambda;\Lambda - \delta\Lambda)
    {}- a_{\rm bare}^t \Pi_\perp^t (\Lambda;\Lambda - \delta\Lambda)
   \Bigr],
\nonumber\\
 &&
 a^s(\Lambda - \delta\Lambda) = a_{\rm bare}^s
 {}+ \frac{1}{F_{\pi,{\rm bare}}^t F_{\pi,{\rm bare}}^s}
\nonumber\\
 &&\times
   \Bigl[ \Pi_\parallel^s (\Lambda;\Lambda - \delta\Lambda)
    {}- a_{\rm bare}^s \Pi_\perp^s (\Lambda;\Lambda - \delta\Lambda)
   \Bigr],
\label{at as}
\end{eqnarray}
where $\Pi_{\perp,\parallel}^{t,s}(\Lambda;\Lambda - \delta\Lambda)$
denotes the quantum correction obtained by integrating the modes out 
between $[\Lambda - \delta\Lambda, \Lambda ]$.
We show the diagrams for contributions to $\Pi_\perp^{\mu\nu}$
and $\Pi_\parallel^{\mu\nu}$ at one-loop level
in Figs.~\ref{fig:AAdiagrams} and~\ref{fig:VVdiagrams}.
The contributions (a) in Fig.~\ref{fig:AAdiagrams} and (a) in
Fig.~\ref{fig:VVdiagrams} are proportional to $g_{L,{\rm bare}}^2$. The
contributions (c) in Fig.~\ref{fig:AAdiagrams} and (d) in
Fig.~\ref{fig:VVdiagrams} are proportional to $(a_{\rm bare}^t-1)$. 
Taking $g_{L,{\rm bare}} = 0$ and $a_{\rm bare}^t = a_{\rm bare}^s =1$, 
these contributions vanish. 
We note that $\sigma$ (i.e., longitudinal vector meson) is massless and
the chiral partner of pion at the critical temperature. 
Then the contributions (b) and (c) in Fig.~\ref{fig:VVdiagrams} have a
symmetry factor $1/2$ respectively and are obviously equal to the
contribution (b) in Fig.~\ref{fig:AAdiagrams}, i.e.,
$\Pi_\perp^{(b)\mu\nu} = \Pi_\parallel^{(b)+(c)\mu\nu}$. 
Thus from Eq.~(\ref{at as}), we obtain
\begin{eqnarray}
 &&
 a^t(\Lambda - \delta\Lambda)
 = a_{\rm bare}^t = 1,
\nonumber\\
 &&
 a^s(\Lambda - \delta\Lambda)
 = a_{\rm bare}^s = 1.
\end{eqnarray}
This implies that $a^t$ and $a^s$ are not renormalized at the scale
$(\Lambda - \delta\Lambda)$. Similarly, we include the corrections
below the scale $(\Lambda - \delta\Lambda)$ in turn, and find that
$a^t$ and $a^s$ do not receive the quantum corrections. Eventually
we conclude that $a^t = a^s =1$ is a fixed point of the RGEs for
$a^t$ and $a^s$.

{}From the above, we find that $(g_L, a^t, a^s) = (0,1,1)$ is 
a fixed point of the combined RGEs for $g_L, a^t$ and $a^s$. 
Thus the VM condition is given by
\begin{eqnarray}
\qquad
 & g_L \to 0,&  \nonumber\\
 & a^t \to 1, \quad
    a^s \to 1&  \quad \mbox{for}\,\,T \to T_c.
\label{evm}
\end{eqnarray}
The vector meson mass is never generated at the critical temperature
since the quantum correction to $M_\rho^2$ is proportional to $g_L^2$.
Because of $g_L \to 0$,
the transverse vector meson at the critical point, in any energy scale, 
decouples from the vector current correlator.
The VM condition for $a^t$ and $a^s$ leads to the equality between
the $\pi$ and $\sigma$ (i.e., longitudinal vector meson) velocities:
\begin{eqnarray}
\qquad
 \bigl( V_\pi / V_\sigma \bigr)^4
 &=& \bigl( F_\pi^s F_\sigma^t / F_\sigma^s F_\pi^t \bigr)^2
\nonumber\\
 &=& a^t / a^s 
 \stackrel{T \to T_c}{\to} 1.
\label{vp=vs}
\end{eqnarray}
This is easily understood from a point of view of the VM
since the longitudinal vector meson becomes the chiral partner
of pion.
We note that this condition
$V_\sigma = V_\pi$ holds independently of the value of the bare
pion velocity which is to be determined through the Wilsonian matching.

\setcounter{equation}{0}
\section{Pion Velocity at Critical Temperature}
\label{PVCT}


As we mentioned in Introduction,
the intrinsic temperature dependence generates the effect of Lorentz
symmetry breaking at bare level.
Then how does the Lorentz non-invariance at bare level influence
physical quantities?
In order to make it clear, in this section
we study the pion decay constants and the pion velocity 
near the critical temperature.

Following Ref.~\cite{HKRS:SUS},
we define the on-shell of the pion from the pole
of the longitudinal component $G_A^L$
of the axial-vector current correlator.
This pole structure is expressed
by temporal and spatial components of the two-point function 
$\Pi_\perp^{\mu\nu}$. 
The temporal and spatial pion decay constants are expressed as follows
~\cite{HKRS:SUS,HS:VVD}:
\begin{eqnarray}
 \qquad
 &\bigl( f_\pi^t (\bar{p};T) \bigr)^2 
  = \Pi_\perp^t(V_\pi\bar{p},\bar{p};T),&
\nonumber\\
 & f_\pi^t (\bar{p};T) f_\pi^s (\bar{p};T) 
  = \Pi_\perp^s(V_\pi\bar{p},\bar{p};T),&
\end{eqnarray}
where the on-shell condition $p_0 \to V_\pi\bar{p}$ was taken.
We divide the two-point function $\Pi_\perp^{\mu\nu}$ into two parts, 
zero temperature (vacuum) and non-zero temperature parts, as $
\Pi_\perp^{\mu\nu} = \Pi_\perp^{(\rm vac)\mu\nu} +
   \bar{\Pi}_\perp^{\mu\nu}$.
The quantum correction is included in the vacuum part
$\Pi_\perp^{(\rm vac)\mu\nu}$, and the hadronic thermal correction
is in $\bar{\Pi}_\perp^{\mu\nu}$. In the present
perturbative analysis, we obtain the pion velocity
as~\cite{HS:VVD}
\begin{eqnarray}
 v_\pi^2 (\bar{p};T)
 &=& \frac{f_\pi^s (\bar{p};T)}
          {f_\pi^t (\bar{p};T)}
\nonumber\\
 &=& V_\pi^2 + \widetilde{\Pi}_\perp (V_\pi\bar{p},\bar{p})
\nonumber\\
&&\hspace*{-1cm}
 {}+ \frac{\bar{\Pi}_\perp^s (V_\pi\bar{p},\bar{p};T) -
    V_\pi^2\bar{\Pi}_\perp^t (V_\pi\bar{p},\bar{p};T)}
   {\bigl( F_\pi^t \bigr)^2}\,,
\label{v_pi-def}
\end{eqnarray}
where $\widetilde{\Pi}_\perp (V_\pi\bar{p},\bar{p})$ denotes
a possible finite renormalization effect.
Note that the renormalization condition on $V_\pi$ is
determined as $\widetilde{\Pi}_\perp (V_\pi\bar{p},\bar{p})|_{\bar{p}=0}=0$.

In the following, we study the quantum and hadronic corrections to
the pion velocity for $T \to T_c$, on the assumption
of the VM conditions~(\ref{evm}). As we defined above, the
two-point function associated with the pion velocity
$v_\pi(\bar{p};T)$ is $\Pi_\perp^{\mu\nu}(p_0,\bar{p};T)$.
The diagrams contributing to $\Pi_\perp^{\mu\nu}$ are shown 
in Fig.~\ref{fig:AAdiagrams}.
As mentioned in
section~\ref{VMFP}, diagram (a) is proportional to $g_L$ and
diagram (c) has the factor $(a^t - 1)$.
Then these contributions vanish at the critical point. 
We consider the contribution from diagram (b) only. 

First we evaluate the quantum correction to the vacuum part
$\Pi_\perp^{(\rm vac)(b)\mu\nu}$.
This is expressed as
\begin{eqnarray}
&&
 \Pi_\perp^{\rm{(vac)}(b)\mu\nu}(p_0,\bar{p})
 = N_f \int\frac{d^n k}{i(2\pi)^n} \nonumber\\
&&\times
   \frac{\Gamma^\mu(k;p) \Gamma^\nu(-k;-p)}
    {[-k_0^2 + V_\pi^2 \bar{k}^2]
     [M_\rho^2 - (k_0 - p_0)^2 + V_\sigma^2 |\vec{k} - \vec{p}|^2]},
\label{Pi-AA-vac-b}
\nonumber\\
\end{eqnarray}
where $\Gamma^\mu$ denotes the ${\cal A}{\pi}{\sigma}$
vertex as
\begin{eqnarray}
&&
 \Gamma^\mu(k;p)
  = \frac{i}{2}\sqrt{a^t}\,g_{\bar{\mu}}^\mu
    [ u^{\bar{\mu}}u^{\bar{\nu}} \nonumber\\
&&\qquad{}+
    V_\sigma^2 (g^{\bar{\mu}\bar{\nu}}-u^{\bar{\mu}}u^{\bar{\nu}}) ]
    (2k-p)_{\bar{\nu}}.
\label{vertex}
\end{eqnarray}
We note that the spatial component of this vertex $\Gamma^i$ has
an extra-factor $V_\sigma^2$ as compared with the temporal one. In
the present analysis it is important to include the quadratic
divergences to obtain the RGEs in the Wilsonian sense. In
Refs.~\cite{Harada:1999zj,HY:WM,HY:PR}, the dimensional
regularization was adopted and the quadratic divergences were
identified with the presence of poles of ultraviolet origin at
$n=2$~\cite{Veltman:1980mj}. 
In this paper when we evaluate four dimensional integral,
we first integrate over $k_0$ from $-\infty$ to $\infty$.
Then we carry out the integral over three-dimensional momentum
$\vec{k}$
with three-dimensional cutoff $\Lambda_3$.
In order to be consistent with ordinary regularization in
four dimension~\cite{Harada:1999zj,HY:WM,HY:PR},
we use the following replacement associated with 
quadratic divergence:
\begin{eqnarray}
&&
 \Lambda_3 \to \frac{1}{\sqrt{2}}\Lambda_4
            = \frac{1}{\sqrt{2}}\Lambda,
\\
 &&\int\frac{d^{n-1} \bar{k}}{(2\pi)^{n-1}}\frac{1}{\bar{k}}
  \to \frac{\Lambda^2}{8\pi^2}, 
\nonumber\\
 &&\int\frac{d^{n-1} \bar{k}}{(2\pi)^{n-1}}
   \frac{\bar{k}^i \bar{k}^j}{\bar{k}^3}
  \to -\delta^{ij}\frac{\Lambda^2}{8\pi^2}.
\end{eqnarray}
When we make these replacement,
the present method of integral preserves chiral symmetry.

As shown in Appendix A, 
$\Pi_\perp^{{\rm (vac)}t}$ and $\Pi_\perp^{{\rm (vac)}s}$ are 
independent of the external momentum.
Accordingly, the finite renrmalization effect $\widetilde{\Pi}_\perp$ 
is also independent of the external momentum and then vanishes:
\begin{equation}
\qquad
 \widetilde{\Pi}_\perp (V_\pi \bar{p},\bar{p}) = 0.
\label{fre}
\end{equation}
Thus in the following,
we take the external momentum as zero.
In that case, the temporal and spatial components of
$\Pi_\perp^{{\rm (vac)}\mu\nu}$ are expressed as
$\Pi_\perp^{\rm (vac)t} = \Pi_\perp^{\rm (vac)00}$ and
$\Pi_\perp^{\rm (vac)s} = - (\delta^{ij}/3)\Pi_\perp^{{\rm (vac)}ij}$.
Taking the VM limit ($M_\rho \to 0$ and $V_\sigma \to V_\pi$),
these components become
\begin{eqnarray}
&&
 \lim_{\rm VM}
 \Pi_\perp^{\rm{(vac)}00}(p_0 = \bar{p} = 0) \nonumber\\
&&
 = \frac{N_f}{4}\int\frac{d k_0 d^{n-1}\bar{k}}{i(2\pi)^n}
   \frac{4k_0^2}{[-k_0^2 + V_\pi^2 \bar{k}^2]^2} \nonumber\\
&&
 = - \frac{N_f}{4}\int\frac{d^{n-1}\bar{k}}{(2\pi)^{n-1}}
     \frac{1}{V_\pi \bar{k}} \nonumber\\
&&
 = -\frac{N_f}{4}\frac{1}{V_\pi}\frac{\Lambda^2}{8\pi^2},
\nonumber\\
&&
 \lim_{\rm VM}
 \Pi_\perp^{\rm{(vac)}ij}(p_0 = \bar{p} = 0) \nonumber\\
&&
 = - \frac{N_f}{4} (V_\pi^2)^2
   \int\frac{d k_0 d^{n-1}\bar{k}}{i(2\pi)^n}
   \frac{4\bar{k}^i \bar{k}^j}{[-k_0^2 + V_\pi^2 \bar{k}^2]^2}
\nonumber\\
&&
 = - \frac{N_f}{4}\, V_\pi^4
   \int\frac{d^{n-1}\bar{k}}{(2\pi)^{n-1}}
   \frac{\bar{k}^i \bar{k}^j}{(V_\pi \bar{k})^3} \nonumber\\
&&
 = \frac{N_f}{4}\,V_\pi\,\delta^{ij}\,\frac{\Lambda^2}{8\pi^2}.
\end{eqnarray}
Thus we obtain the temporal and spatial parts as
\begin{eqnarray}
&&
 \lim_{\rm VM}
 \Pi_\perp^{\rm{(vac)}t}(p_0 = \bar{p} = 0) 
 = -\frac{N_f}{4}\frac{1}{V_\pi}\frac{\Lambda^2}{8\pi^2},
\nonumber\\
&&
 \lim_{\rm VM}
 \Pi_\perp^{\rm{(vac)}s}(p_0 = \bar{p} = 0) 
 = -\frac{N_f}{4}\,V_\pi\,\frac{\Lambda^2}{8\pi^2}.
\label{qc}
\end{eqnarray}
These quadratic divergences are renormalized by
$(F_{\pi,{\rm bare}}^t)^2$ and
$F_{\pi,{\rm bare}}^t F_{\pi,{\rm bare}}^s$, respectively.
Then RGEs for the parameters $(F_\pi^t)^2$ and
$F_\pi^t F_\pi^s$ are expressed as
\begin{eqnarray}
&&
 \mu \frac{d\bigl( F_\pi^t \bigr)^2}{d\mu}
 = \frac{N_f}{(4\pi)^2}\frac{1}{V_\pi}\mu^2,
\label{rge-tt}\\
&&
 \mu \frac{d\bigl( F_\pi^t F_\pi^s \bigr)}{d\mu}
 = \frac{N_f}{(4\pi)^2}V_\pi\,\mu^2.
\label{rge-ts}
\end{eqnarray}
Both $(F_\pi^t)^2$ and $F_\pi^t F_\pi^s$ scale following the quadratic
running $\mu^2$.
However, the coefficient of $\mu^2$ in the RGE for $(F_\pi^t)^2$
is different from that for $F_\pi^t F_\pi^s$.

When we use these RGEs, the scale dependence of the parametric 
pion velocity is
\begin{eqnarray}
 \mu \frac{dV_\pi^2}{d\mu}
 &=& \mu \frac{d\bigl( F_\pi^t F_\pi^s / (F_\pi^t)^2 \bigr)}{d\mu}
 \nonumber\\
 &=& \frac{1}{\bigl(F_\pi^t\bigr)^4}\frac{N_f}{(4\pi)^2}
     \Bigl[ V_\pi \bigl(F_\pi^t\bigr)^2 - F_\pi^t F_\pi^s
      \frac{1}{V_\pi} \Bigr]\mu^2 \nonumber\\
 &=& 0\,. \label{v-scale}
\end{eqnarray}
This implies that {\it the parametric pion velocity at the critical
temperature does not scale.}
In other words, the Lorentz non-invariance at bare level is not
enhanced through the RGEs as long as we consider the pion velocity.
As we noted below Eq.~(\ref{vertex}),
the factor $V_\sigma^2$ is in the spatial component of the vertex
$\Gamma^\mu$.
If $V_\sigma$ were not equal to $V_\pi$,
the coefficients of running in the right-hand-side of
Eqs.~(\ref{rge-tt}) and (\ref{rge-ts}) would change.
However, since the VM conditions do guarantee $V_\sigma = V_\pi$,
the quadratic running caused from $\Lambda^2$ in $(F_\pi^t)^2$
and $F_\pi^t F_\pi^s$ are exactly canceled in the second line of
Eq.~(\ref{v-scale}).

Next we study the hadronic thermal correction to the pion velocity 
at the critical temperature. The temporal and spatial parts of the
hadronic thermal correction $\bar{\Pi}_\perp^{\mu\nu}$ contribute
to the pion velocity, which have the same structure as those of
the quantum correction $\Pi_\perp^{(\rm vac)\mu\nu}$, except for a
Bose-Einstein distribution function. Thus by the replacement of
$\Lambda^2 / (4\pi)^2$ with $T^2 / 12$ in $\Pi_\perp^{\rm
(vac)t,s}$, hadronic corrections to the temporal and spatial parts
of $\bar{\Pi}_\perp^{\mu\nu}$ are obtained as follows:
\begin{eqnarray}
&&
 \lim_{\rm VM}
 \bar{\Pi}_\perp^{t}(p_0,\bar{p};T)
 = -\frac{N_f}{24}\frac{1}{V_\pi}T_c^2,
\nonumber\\
&&
 \lim_{\rm VM}
 \bar{\Pi}_\perp^{s}(p_0,\bar{p};T)
 = -\frac{N_f}{24}\,V_\pi\,T_c^2.
\label{hc}
\end{eqnarray}

Substituting Eq.~(\ref{hc}) into Eq.~(\ref{v_pi-def})
with $\widetilde{\Pi}_\perp = 0$ as shown in Eq.~(\ref{fre}),
we obtain the physical pion velocity in the VM as
\begin{eqnarray}
&&
 v_{\pi}^2(\bar{p};T) \nonumber\\
&&
 \stackrel{T \to T_c}{\to}
    V_\pi^2
   {}+ \frac{\bar{\Pi}_\perp^s (V_\pi\bar{p},\bar{p};T_c) -
    V_\pi^2\bar{\Pi}_\perp^t (V_\pi\bar{p},\bar{p};T_c)}
   {\bigl( F_\pi^t \bigr)^2}
\nonumber\\
&&\,\,
   = V_\pi^2.
\label{velocity}
\end{eqnarray}
Since the parametric pion velocity in the VM does not scale
with energy [see Eq.~(\ref{v-scale})],
$V_\pi$ in the above expression is equivalent
to the bare pion velocity:
\begin{equation}
\qquad
 v_\pi(\bar{p};T) = V_{\pi,{\rm bare}}(T)
 \qquad \mbox{for}\, T \to T_c.
\label{phys=bare}
\end{equation}
This implies that {\it the pion velocity in the limit $T \to T_c$
receives neither hadronic nor quantum corrections due
to the protection by the VM.} This is our main result.

In order to clarify the reason of this non-renormalization property,
let us recall the fact that only the diagram (b) 
in Fig.~\ref{fig:AAdiagrams} contributes to the pion velocity 
at the critical temperature.
Away from the critical temperature, the contribution of the massive 
$\sigma$ (i.e., the longitudinal mode of massive vector meson) is 
suppressed owing to the Boltzmann factor $\exp [-M_\rho / T]$, 
and then only the pion loop contributes to the pion velocity.
Then there exists the ${\cal O}(T^4)$ correction to the pion velocity
~\cite{HS:VVD}.
Near the critical temperature, on the other hand,
$\sigma$ becomes massless due to the VM
since $\sigma$ (i.e., the longitudinal vector meson) becomes 
the chiral partner of the pion.
Thus the absence of the hadronic corrections in the pion velocity
at the critical temperature is due to 
the exact cancellation between the contribution of pion and that 
of its chiral partner $\sigma$.
Similarly the quantum correction generated from the pion loop 
is exactly cancelled by that from the $\sigma$ loop.


\setcounter{equation}{0}
\section{Summary and Discussions}
\label{SD}


In this paper, we started from the Lorentz non-invariant HLS
Lagrangian at bare level and studied the pion velocity at the
critical temperature based on the VM. From the analysis of the
quantum and hadronic thermal corrections to the pion velocity, we obtained
the result that {\it the pion velocity in the limit $T \to T_c$
is equal to the bare pion velocity.} In other words, the pion
velocity does not receive either quantum or hadronic corrections
at the critical temperature.
This occurs due to the exact cancellation between the contribution of
pion and that of the longitudinal vector meson (i.e., the chiral partner
of pion).

Now we consider the meaning of our result~(\ref{phys=bare}). Based
on the point of view that the bare HLS theory is defined from QCD,
we presented the VM conditions realizing the chiral symmetry in
QCD consistently, i.e, $(g_L,a^t,a^s) \to (0,1,1)$ for $T \to
T_c$. This is the fixed point of the RGEs for the parameters
$g_L, a^t$ and $a^s$. 
As we showed in section~\ref{PVCT}, 
although both pion decay constants $(F_\pi^t)^2$ 
and $F_\pi^t F_\pi^s$ scale following the quadratic running, 
$(F_\pi^t)^2$ and $F_\pi^t F_\pi^s$ show a different running 
since the coefficient of $\mu^2$ in Eq.~(\ref{rge-tt}) is
different from that in Eq.~(\ref{rge-ts}).
Nevertheless in the
pion velocity at the critical temperature, the quadratic running
in $(F_\pi^t)^2$ is exactly cancelled by that in $F_\pi^t F_\pi^s$
[see second line of Eq.~(\ref{v-scale})]. There it was crucial for
intricate cancellation of the quadratic running that the velocity
of $\sigma$ (i.e., longitudinal vector meson) is equal to its
chiral partner, i.e., $V_\sigma \to V_\pi$ for $T \to T_c$. Note
that this is not an extra condition but a consequence from the VM
conditions for $a^t$ and $a^s$; we started simply from the VM
conditions alone and found that $V_\pi$ does not receive quantum
corrections at the restoration point. 
As we showed in Eq.~(\ref{hc}), the
hadronic correction to $(F_\pi^t)^2$ is different from that to
$F_\pi^t F_\pi^s$. 
In the pion velocity, however,  
the hadronic correction from $(F_\pi^t)^2$ is exactly
cancelled by that from $F_\pi^t F_\pi^s$ [see second line of
Eq.~(\ref{velocity})]. The VM conditions guarantee these exact
cancellations of the quantum and hadronic corrections. 
This implies that $(g_L,a^t,a^s,V_\pi) = (0,1,1,\mbox{any})$ forms
a fixed line for four RGEs of $g_L, a^t, a^s$ and $V_\pi$.
When one point on this fixed line is selected through the matching
procedure as done in Ref.~\cite{HKRS:pv},
namely the value of $V_{\pi,{\rm bare}}$ is fixed,
the present result implies that the point does not move in a subspace
of the parameters. 
This is likely the manifestation of a new fixed point 
in the Lorentz non-invariant formulation of the VM. 
Approaching the restoration point of chiral symmetry, 
the physical pion velocity itself would flow into the fixed point.

Several comments are in order:

We should distinguish the consequences within HLS/VM from those
beyond HLS/VM.
Clearly the determination of the definite value of
the bare pion velocity is done outside HLS/VM.
On the other hand, our main result~(\ref{phys=bare}) holds
independently of the value of the bare pion velocity itself.
Applying this result to the case
where one starts from the bare HLS theory with Lorentz invariance,
i.e., $V_{\pi,{\rm bare}}=1$,
one finds that the pion velocity at $T_c$ becomes the speed of light since
$v_\pi = V_{\pi,{\rm bare}} = 1$,
as obtained in Ref.~\cite{HKRS:SUS}.

As a consequence of the relation~(\ref{phys=bare}),
we can determine the temporal and spatial pion decay constants
at the critical temperature when we take the bare pion velocity
as finite.
In the following, we study these decay constants
and discuss their determinations based on Eq.~(\ref{phys=bare}).
Using Eq.~(\ref{v-scale}), we solve the RGEs~(\ref{rge-tt})
and (\ref{rge-ts}) and
obtain a relation between two parametric pion decay constants as
$F_\pi^t(0;T_c)F_\pi^s(0;T_c) = V_\pi^2 \bigl( F_\pi^t(0;T) \bigr)^2$.
From this and (\ref{hc}),
the temporal and spatial pion decay constants with the quantum and
hadronic corrections are obtained as
\begin{eqnarray}
\quad
 \bigl( f_\pi^t \bigr)^2
 &=& \Bigl( F_\pi^t(0;T_c)\Bigr)^2 -
\frac{N_f}{24}\frac{1}{V_\pi}T_c^2,
\nonumber\\
 f_\pi^t\,f_\pi^s
 &=& F_\pi^t(0;T_c)F_\pi^s(0;T_c) - \frac{N_f}{24}V_\pi\,T_c^2
\nonumber\\
 &=& V_\pi^2 \bigl( f_\pi^t \bigr)^2.
\end{eqnarray}
Since the order parameter $(f_\pi^t f_\pi^s)$ vanishes as expected
at the critical temperature, we find that $ f_\pi^t f_\pi^s =
V_\pi^2 \bigl( f_\pi^t \bigr)^2 = 0\,$. Multiplying both side by
$v_\pi^2 = V_\pi^2$, the above expression is reduced to
\begin{equation}
\qquad
 (f_\pi^s)^2 = V_\pi^4 (f_\pi^t)^2 = 0.
\label{s-t}
\end{equation}
Now, the spatial pion decay constant vanishes at the critical
temperature, $ f_\pi^s (T_c) = 0\,$. In the case of a vanishing
pion velocity, $f_\pi^t$ can be finite at the restoration point.
On the other hand, when $V_\pi$ is finite, Eq.~(\ref{s-t}) leads
to $f_\pi^t(T_c)=0$. Thus we find that both temporal and spatial
pion decay constants vanish simultaneously at the critical
temperature when the bare pion velocity is determined as finite.

In order to know the value of the (bare) pion velocity, we need to
specify a method that determines the bare parameters of the
effective field theory. As we stressed in subsection~\ref{VMC}, the
{\it bare} parameters of the HLS Lagrangian are determined by the
underlying QCD. One possible way to determine them is the
Wilsonian matching proposed in Ref.~\cite{HY:WM} which is done by
matching the axial-vector and vector current correlators derived
from the HLS with those by the operator product expansion (OPE) in
QCD at the matching scale $\Lambda$. From the analysis performed
on the basis of a Wilsonian matching, the bare pion velocity at
the critical temperature is found to be finite, i.e.,
$V_{\pi,{\rm bare}} \neq 0$~\cite{HKRS:pv}.
Thus, by combining
Eq.~(\ref{phys=bare}) with estimation of $V_{\pi,{\rm bare}}$,
the value of the physical pion velocity $v_\pi(T)$ at
the critical temperature is obtained to be finite~\cite{HKRS:pv}.
This is in contrast to the result obtaied from the chiral theory
~\cite{Son:2001ff,Son:2002ci},
where the relevent degree of freedom near $T_c$ is only the pion.
Their result is that the pion velocity becomes zero for $T \to T_c$.
Therefore from the experimental data,
we may be able to distinguish which picture is correct, 
$v_\pi \sim 1$ or $v_\pi \to 0$.

\section*{Acknowledgment}


The author would like to thank
Professor Masayasu Harada, Doctor Youngman Kim and
Professor Mannque Rho
for many useful discussions and comments.

\appendix

\begin{flushleft}
 {\bf {Appendix}}
\end{flushleft}

\setcounter{section}{0}
\renewcommand{\thesection}{\Alph{section}}
\setcounter{equation}{0}
\renewcommand{\theequation}{\Alph{section}.\arabic{equation}}

\section{Two-point Function}

In this appendix, we show that the temporal and spatial components
of the two-point function $\Pi_\perp^{{\rm (vac)}\mu\nu}$ are
independent of the external momentum in the VM limit.

{}From Eq.~(\ref{Pi-AA-vac-b}),
$\Pi_\perp^{{\rm (vac)(b)}\mu\nu}$ is rewritten as
\begin{eqnarray}
&&
 \Pi_\perp^{\rm{(vac)(b)}\mu\nu}(p_0,\bar{p})
\nonumber\\
&&
 = N_f\,\frac{a^t}{4} X_{\bar{\mu}}^\mu X_{\bar{\nu}}^\nu
     B^{\rm{(vac)}\bar{\mu}\bar{\nu}}(p_0,\bar{p};M_\rho,0) \nonumber\\
&&
 \equiv N_f\,\frac{a^t}{4} 
     \tilde{B}^{\rm{(vac)}\mu\nu}(p_0,\bar{p};M_\rho,0),
\end{eqnarray}
where we define
\begin{equation}
\qquad
 X_{\bar{\mu}}^\mu
 = u^\mu u_{\bar{\mu}} + V_\sigma^2 (g_{\bar{\mu}}^\mu - 
   u^\mu u_{\bar{\mu}}).
\end{equation}
In the above expression,
we define the function $B^{(\rm vac)\mu\nu}$ contributed to 
the diagram (b) in Fig.~\ref{fig:AAdiagrams} as
\begin{eqnarray}
 &&B^{\rm{(vac)}\mu\nu}(p_0,\bar{p};M_\rho,0)
 = \int\frac{dk_0}{i(2\pi)}\int\frac{d^3\bar{k}}{(2\pi)^3}
\nonumber\\
&&\quad\times
   \frac{(2k-p)^\mu (2k-p)^\nu}{[k_0^2 - \omega_\pi^2]
    [(k_0 - p_0)^2 - (\omega_\rho^p)^2]},
\end{eqnarray}
with
\begin{eqnarray}
\qquad\quad
 \omega_\pi^2 &=& V_\pi^2 \bar{k}^2, \nonumber\\
 (\omega_\rho^p)^2 &=& V_\sigma^2 |\vec{k}-\vec{p}|^2 + M_\rho^2.
\end{eqnarray}
In terms of each components of $B^{(\rm vac)\mu\nu}$,
the temporal and spatial parts of $\tilde{B}^{\mu\nu}$ are given by
\begin{eqnarray}
 &&\tilde{B}^{\rm{(vac)}t}(p_0,\bar{p};M_\rho,0)
   = \Bigl[ B^{\rm{(vac)}t}(p_0,\bar{p};M_\rho,0) 
\nonumber\\
&&\quad{}+
     (1-V_\sigma^2)\frac{\bar{p}^2}{p^2}
      B^{\rm{(vac)}L}(p_0,\bar{p};M_\rho,0)\Bigr], \nonumber\\
 &&\tilde{B}^{\rm{(vac)}s}(p_0,\bar{p};M_\rho,0)
   = V_\sigma^4 \Bigl[ B^{\rm{(vac)}s}(p_0,\bar{p};M_\rho,0) 
\nonumber\\
&&\quad{}+
     \frac{1-V_\sigma^2}{V_\sigma^2}\,\frac{p_0^2}{p^2}
      B^{\rm(vac)L}(p_0,\bar{p};M_\rho,0)\Bigr].
\label{ef}
\end{eqnarray}
By using the expressions in Ref.~\cite{HS:VVD},
the componets $B^{\rm (vac)t}, B^{\rm (vac)s}$ and $B^{\rm (vac)L}$ 
take the following forms:
\begin{eqnarray}
 &&
 B^{\rm (vac)t}(p_0,\bar{p};M_\rho,0)
\nonumber\\
&&
  = \int\,\frac{d^3 k}{(2\pi)^3}
   \Biggl[ \frac{-1}{2\omega_\pi}
    \frac{(2\omega_\pi - p_0)^2}
         {(\omega_\pi - p_0)^2 - (\omega_\rho^p)^2}
\nonumber\\
&&\qquad\qquad\qquad\qquad{}+
    \frac{-1}{2\omega_\rho^p}
    \frac{(2\omega_\rho^p + p_0)^2}
         {(\omega_\rho^p + p_0)^2 - \omega_\pi^2} \nonumber\\
 &&\quad
  {}-\frac{\vec{p}\cdot(2\vec{k}-\vec{p})}{p_0}
   \Biggl\{ \frac{1}{2\omega_\pi}
     \frac{2\omega_\pi - p_0}
          {(\omega_\pi - p_0)^2 - (\omega_\rho^p)^2} 
\nonumber\\
&&\qquad\qquad\qquad\qquad{}+
     \frac{1}{2\omega_\rho^p}
     \frac{2\omega_\rho^p + p_0}
          {(\omega_\rho^p + p_0)^2 - \omega_\pi^2}
   \Biggr\}
  \Biggr],
\nonumber\\
 &&
 B^{\rm (vac)s}(p_0,\bar{p};M_\rho,0) \nonumber\\
 &&
  = \int\,\frac{d^3 k}{(2\pi)^3}
   \Biggl[ \frac{(2\vec{k}\cdot\vec{p}-\bar{p}^2)^2}{\bar{p}^2}
\nonumber\\
&&\times
    \Biggl\{ 
     \frac{1}{2\omega_\pi}
     \frac{1}{(\omega_\pi - p_0)^2 - (\omega_\rho^p)^2} 
\nonumber\\
&&\qquad\qquad\qquad{}+
     \frac{1}{2\omega_\rho^p}
     \frac{1}{(\omega_\rho^p + p_0)^2 - \omega_\pi^2} 
    \Biggr\} \nonumber\\
 &&
  {}+\frac{p_0\vec{p}\cdot (2\vec{k}-\vec{p})}{\bar{p}^2}
\nonumber\\
&&\times
   \Biggl\{ \frac{1}{2\omega_\pi}
     \frac{2\omega_\pi - p_0}
          {(\omega_\pi - p_0)^2 - (\omega_\rho^p)^2} 
\nonumber\\
&&\qquad\qquad\qquad{}+
     \frac{1}{2\omega_\rho^p}
     \frac{2\omega_\rho^p + p_0}
          {(\omega_\rho^p + p_0)^2 - \omega_\pi^2}
   \Biggr\}
  \Biggr],
\nonumber\\
 &&
 B^{\rm (vac)L}(p_0,\bar{p};M_\rho,0) \nonumber\\
 &&
  = \int\,\frac{d^3 k}{(2\pi)^3}
  \frac{p^2 \vec{p}\cdot (2\vec{k}-\vec{p})}{p_0 \bar{p}^2}
\nonumber\\
&&\qquad\times
  \Biggl[ 
     \frac{1}{2\omega_\pi}
     \frac{2\omega_\pi - p_0}
          {(\omega_\pi - p_0)^2 - (\omega_\rho^p)^2} 
\nonumber\\
&&\qquad\qquad\qquad{}+
     \frac{1}{2\omega_\rho^p}
     \frac{2\omega_\rho^p + p_0}
          {(\omega_\rho^p + p_0)^2 - \omega_\pi^2}
  \Biggr].
\end{eqnarray}

Now we take the VM limit.
Note that the VM condition for $a^t$ and $a^s$ implies 
that $\sigma$ velocity becomes equal to the pion velocity, 
$V_\sigma \to V_\pi$ for $T \to T_c$ as shown in Eq.~(\ref{vp=vs}).
The components $B^{\rm{(vac)}t}, B^{\rm{(vac)}s}$ and $B^{\rm{(vac)}L}$
are calculated as follows:
\begin{eqnarray}
 &&\lim_{\rm VM}B^{\rm{(vac)}t}(p_0,\bar{p};M_\rho,0)
   = \int\frac{d^3 k}{(2\pi)^3} \frac{-1}{\omega_\pi}
\nonumber\\
&&\times
     \frac{{\cal I}(\bar{k};p_0,\bar{p}) + \vec{p}\cdot(2\vec{k}-\vec{p})
      {\cal J}(\bar{k};p_0,\bar{p})}
      {[(\omega_\pi - p_0)^2 - (\omega_\pi^p)^2]
       [(\omega_\pi + p_0)^2 - (\omega_\pi^p)^2]}, \label{t}
\nonumber\\
{}\\
 &&\lim_{\rm VM}B^{\rm{(vac)}s}(p_0,\bar{p};M_\rho,0)
   = \int\frac{d^3 k}{(2\pi)^3} \frac{1}{\omega_\pi}
     \frac{1}{\bar{p}^2}
\nonumber\\
&&\times
     \frac{1}
      {[(\omega_\pi - p_0)^2 - (\omega_\pi^p)^2]
       [(\omega_\pi + p_0)^2 - (\omega_\pi^p)^2]}
\nonumber\\
&&\times
 \Bigl[
 \Bigl( \vec{p}\cdot(2\vec{k}-\vec{p})\bigr)^2 
      {\cal K}(\bar{k};p_0,\bar{p}) 
\nonumber\\
&&\qquad\qquad{}+ 
      p_0^2 \vec{p}\cdot(2\vec{k}-\vec{p})
      {\cal J}(\bar{k};p_0,\bar{p})
 \Bigr], \label{s}\\
 &&\lim_{\rm VM}B^{\rm{(vac)}L}(p_0,\bar{p};M_\rho,0)
   = \int\frac{d^3 k}{(2\pi)^3} 
     \frac{1}{\omega_\pi}
     \frac{p^2}{\bar{p}^2}
\nonumber\\
&&\times
     \frac{\vec{p}\cdot(2\vec{k}-\vec{p}) {\cal J}(\bar{k};p_0,\bar{p})}
      {[(\omega_\pi - p_0)^2 - (\omega_\pi^p)^2]
       [(\omega_\pi + p_0)^2 - (\omega_\pi^p)^2]}, \label{L}
\nonumber\\
{}
\end{eqnarray}
where we define the functions as
\begin{eqnarray}
&&
 {\cal I}(\bar{k};p_0,\bar{p})
\nonumber\\
&&\quad
 = \omega_\pi^2 [ 4\omega_\pi^2 - 4(\omega_\pi^p)^2 - 3p_0^2 ] +
     p_0^2[ p_0^2 - (\omega_\pi^p)^2 ], \nonumber\\
&&
 {\cal J}(\bar{k};p_0,\bar{p})
 = -p_0[ -3\omega_\pi^2 + p_0^2 - (\omega_\pi^p)^2 ], \nonumber\\
&&
 {\cal K}(\bar{k};p_0,\bar{p})
 = \omega_\pi^2 + p_0^2 - (\omega_\pi^p)^2.
\end{eqnarray}
Using Eqs.~(\ref{t})-(\ref{L}) with these functions,
we obtain $\tilde{B}^{\rm{(vac)}t,s}$ as
\begin{eqnarray}
 &&\lim_{\rm VM}\tilde{B}^{\rm{(vac)}t}(p_0,\bar{p};M_\rho,0)
   = \int\frac{d^3 k}{(2\pi)^3} \frac{-1}{\omega_\pi}
\nonumber\\
&&\times
     \frac{{\cal I}(\bar{k};p_0,\bar{p}) + 
       V_\pi^2\vec{p}\cdot(2\vec{k}-\vec{p})
      {\cal J}(\bar{k};p_0,\bar{p})}
      {[(\omega_\pi - p_0)^2 - (\omega_\pi^p)^2]
       [(\omega_\pi + p_0)^2 - (\omega_\pi^p)^2]}, \label{tilde-t}
\nonumber\\
{}\\
 &&\lim_{\rm VM}\tilde{B}^{\rm{(vac)}s}(p_0,\bar{p};M_\rho,0)
   = \int\frac{d^3 k}{(2\pi)^3} \frac{1}{\omega_\pi}
     \frac{1}{V_\pi^2 \bar{p}^2}
\nonumber\\
&&\times
     \frac{1}
      {[(\omega_\pi - p_0)^2 - (\omega_\pi^p)^2]
       [(\omega_\pi + p_0)^2 - (\omega_\pi^p)^2]}
\nonumber\\
&&\times
  \Bigl[
  \Bigl( V_\pi^2 \vec{p}\cdot(2\vec{k}-\vec{p}) \Bigr)^2 
      {\cal K}(\bar{k};p_0,\bar{p}) 
\nonumber\\
&&\qquad{}+ 
      p_0^2 V_\pi^2 \vec{p}\cdot(2\vec{k}-\vec{p})
      {\cal J}(\bar{k};p_0,\bar{p})
  \Bigr]. \label{tilde-s}
\end{eqnarray}
The integrand of the functions $\tilde{B}^{\rm (vac)t}$ and 
$\tilde{B}^{\rm (vac)s}$ in the VM limit are same as
those of $B^{\rm (vac)t}$ and $B^{\rm (vac)s}$ with $V_\pi = 1$
when we make the following replacement in $\tilde{B}^{\rm (vac)t,s}$:
\begin{equation}
\qquad
 V_\pi \bar{k} \to \bar{k}, \qquad
 V_\pi |\vec{k} - \vec{p}| \to |\vec{k} - \vec{p}|.
\label{replacement}
\end{equation}
The functions $B^{\rm (vac)t,s}$ with $M_\rho \to 0$ are
independent of the external momentum $p_0$ and $\bar{p}$~\cite{HS:VVD}
~\footnote{
 In Ref.~\cite{HKRS:SUS} where $V_\pi = 1$ was taken,
 it was shown that
 the hadronic corrections $\bar{\Pi}_\perp^{t,s}(p_0,\bar{p};T)$
 at the VM limit are independent of the external momentum $p_0$
 and $\bar{p}$.
 The structure of the integrand in the vacuum part is the same as
 that in the hadronic part except for the absence
 of the Bose-Einstein distribution function.
 Thus the vacuum part is also independent of $p_0$
 and $\bar{p}$.
}.
Thus
we find that the functions $\tilde{B}^{(\rm vac)t}$ and 
$\tilde{B}^{(\rm vac)s}$ in the VM limit are 
obtained independently of the external momentum $p_0$ and $\bar{p}$:
\begin{eqnarray}
&&
 \lim_{\rm VM}\tilde{B}^{\rm{(vac)}t}(p_0,\bar{p};M_\rho,0)
 = -\frac{1}{V_\pi}\frac{\Lambda^2}{8\pi^2},
\nonumber\\
&&
 \lim_{\rm VM}\tilde{B}^{\rm{(vac)}s}(p_0,\bar{p};M_\rho,0)
 = -{V_\pi}\frac{\Lambda^2}{8\pi^2}.
\end{eqnarray}

\end{document}